\documentclass[12pt]{cernlik}
\usepackage{epsfig}
\usepackage{amssymb}

\def\bea{\begin{eqnarray}}
\def\eea{\end{eqnarray}}

\def\be{\begin{equation}}
\def\ee{\end{equation}}

\begin{document}
\title{Search for correlation between GRB's detected by BeppoSAX \\
and  gravitational wave detectors EXPLORER and NAUTILUS}
\author{
P. Astone$^1$, M. Bassan$^2$, P. Bonifazi$^3$, P. Carelli$^4$,
 G. Castellano$^5$\\ E. Coccia$^2$, C. Cosmelli$^6$, G. D'Agostini$^6$,
S. D'Antonio$^2$,
V. Fafone$^7$, G. Federici$^1$\\ F. Frontera$^8$,
 C. Guidorzi$^9$, A. Marini$^7$,
Y. Minenkov$^2$, I. Modena$^2$\\ G. Modestino$^7$,
A. Moleti$^2$, E. Montanari$^{10}$, G. V. Pallottino$^6$\\ G. Pizzella$^{11}$,
L. Quintieri$^7$, A. Rocchi$^{2}$, F. Ronga$^7$, R. Terenzi$^{12}$,
G. Torrioli$^5$, M. Visco$^{13}$
}
{\it ${}^{1)}$ Istituto Nazionale di Fisica Nucleare INFN, Rome}\\
{\it ${}^{2)}$ University of Rome "Tor Vergata" and INFN, Rome 2}\\
{\it ${}^{3)}$ IFSI-CNR and INFN, Rome}\\
{\it ${}^{4)}$ University of L'Aquila and INFN, Rome 2}\\
{\it ${}^{5)}$ IESS-CNR and INFN, Rome}\\
{\it ${}^{6)}$ University of Rome "La Sapienza" and INFN, Rome}\\
{\it ${}^{7)}$ Istituto Nazionale di Fisica Nucleare INFN, Frascati}\\
{\it ${}^{8)}$ University of Ferrara and IASF-CNR, Bologna}\\
{\it ${}^{9)}$ University of Ferrara, Ferrara}\\
{\it ${}^{10)}$ University of Ferrara, Ferrara and ITA ``I. Calvi'',
Finale Emilia, Modena}\\
{\it ${}^{11)}$ University of Rome "Tor Vergata" and INFN, Frascati}\\
{\it ${}^{12)}$ IFSI-CNR and  INFN, Rome 2}\\
{\it ${}^{13)}$ IFSI-CNR and  INFN, Frascati}\\
\\ \vspace{2.5cm}
\maketitle
\begin{abstract}
Data obtained during five months of 2001
with the gravitational wave (GW) detectors EXPLORER and NAUTILUS
were studied in correlation with the gamma ray burst data (GRB)
obtained with the BeppoSAX satellite.
During this period BeppoSAX was the only GRB satellite in operation, while
EXPLORER and NAUTILUS were the only GW detectors in operation.
 No correlation between the GW data and the GRB bursts was found. 
The analysis, performed over 47 GRB's, 
excludes the presence of signals of amplitude
  $h\ge 1.2 \times 10^{-18}$, with 95\% probability,
   if we allow a time delay between GW bursts and GRB
  within $\pm400$  s, and $h \ge 6.5 \times 10^{-19}$, if the time delay is
  within $\pm5$ s.
The result is also provided in form of scaled likelihood for 
unbiased interpretation and easier use for further analysis.
\end{abstract}

\newpage
\section{Introduction}
\noindent
One of the most important astrophysical phenomena
still lacking an explanation
is the origin of the celestial gamma-ray
bursts (GRB). These are powerful flashes of gamma-rays lasting from less
than one second to tens of seconds, with isotropic distribution in the sky.
They  are observed above the terrestrial atmosphere with X-- gamma--ray
detectors aboard satellites \cite{fi,fis}. Thanks to the
BeppoSAX satellite \cite{Boella}, afterglow emission at lower
wavelengths has been discovered \cite{Costa,Jvp,Frail} and we now know
that at least long ($>1 \,$s) GRB's are at cosmological distances, with
measured red shifts up to 4.5 (see, e.g., review by Djorgovski
\cite{Djorgovski} and references therein).
Among the possible explanations of these events, which involve huge
energy releases (up to $10^{54}$ erg, assuming isotropic emission),
the most likely candidates are the collapse of a very massive star (hypernova)
and the  coalescence of one compact binary system
(see, e.g., reviews by Piran \cite{Piran99} and M\'esz\'aros \cite{Meszaros01}
and references therein). In both cases
emission of gravitational waves (GW) is expected to be associated with them
(e.g. Ref. \cite{ks}).
According to several models, 
the duration of a GW burst is
predicted to be of the order of a few milliseconds for a variety of sources,
including the coalescing and merging black holes and/or neutron star binaries.
Therefore GW bursts can be detected by
the present resonant detectors, designed to detect GW through the excitation
of the quadrupole modes of massive cylinders, resonating at frequencies
near 1 kHz.

At the distances of the GRB sources ($\approx 1$ Gpc), the GW burst
associated with a total conversion of 1-2 solar masses should have amplitude
of the order of $h  \approx 3 \times 10^{-22}$. The present
sensitivity for 1 ms GW pulses of the best GW
antennas with signal to noise ratio (SNR) equal to unity is
$h \approx 4 \times 10^{-19}$ (see e.g. Ref. \cite{piaamaldi}), which requires a
total conversion of one million solar masses at 1 Gpc. However, although detection 
of a gravitational signal associated with a single GRB appears hopeless, 
detection of a signal associated with the sum
of many events could be more realistic. Thus we launched a program devoted
to studying the presence of correlations between GRB events detected
with BeppoSAX and the output signals from gravitational antennas
NAUTILUS and EXPLORER.

Searching for correlation between GRB and GW signals means dealing
with the difference between the emission times for the two types of phenomena.
Furthermore, there is also the fact to consider that the time difference
can vary from burst to burst.
In the present analysis we use an algorithm based on cross-correlating
the outputs of two GW detectors (see \cite{finn,arturo}), thus coping
with the problem of the unknown possible time difference between GRB 
and GW bursts, and also of the unmodelled noise.

\section{Experimental data}
\noindent
The Rome group operates two resonant bar detectors:
EXPLORER \cite{as2}, since 1990, at the CERN laboratories, and
NAUTILUS \cite{nautilus}, since 1995, at the INFN laboratories
in Frascati.
\begin{table}
\centering
\caption{
Main characteristics of the two detectors. $f$ indicates, for each detector,
 the two resonant frequencies and $\Delta f$ indicates the bandwidth.
The relatively larger bandwidth of EXPLORER is due to an improved readout system.
}
\vskip 0.1 in
\begin{tabular}{|c|c|c|c|c|c|c|c|}
\hline
detector&latitude&longitude&orientation&mass&$f$&$T$&$\Delta f$\\
&&&&kg&Hz&K&Hz\\
\hline
EXPLORER&$46.45^o$ N&$6.20^o$ E&$39^o$ E&2270&904.7&2.6&$\approx 9$\\
&&&&&921.3&&\\
NAUTILUS&$41.82^o$ N&$12.67^o$ E&$44^o$ E&2270&906.97&1.5& $\approx 0.4$\\
&&&&&922.46&&\\
\hline
\end{tabular}
\label{dire}
\end{table}
The two detectors, oriented nearly parallel, are very similar.
They have operated over the
last few years with various levels of sensitivity, and since March $1^{st}$
2001 they have been in operation simultaneously
 with the best ever reached sensitivity for millisecond bursts,
of the order of $h \approx (4-5)\times10^{-19}$.
The detectors consist of massive cylindrical bars 3 m long made of
high quality factor Aluminum alloy 5056.
The GW excites the first longitudinal mode of the bar, which is cooled
to liquid helium temperature to reduce the thermal noise. To measure
the bar strain induced by a GW, a secondary mechanical oscillator
tuned to the cited mode is mounted on one bar face and a sensor measures
the displacement of the secondary oscillator.

The data have a sampling time of 4.544 ms and are processed
with a filter matched to delta-like signals 
for the detection of short bursts~\cite{fast} . The filter is adaptive and
makes use of power spectra obtained during periods of two hours.
The filtered output is squared and normalized using the detector calibration,
such that its square gives the energy innovation $x(t)$
for each sample.
In the presence only of well behaved noise due to the thermal
motion of the bar and to the electronic noise of the amplifier,
the probability density function of $x(t)$ is
\be
f(x)=\frac{1}{\sqrt{2\pi\, T_{eff}\,x}}
\exp{\left[{-\frac{x}{2\, T_{eff}}} \right] }
\label{normal}
\ee
where $x(t)$ is  expressed in kelvin units, and
 the average value of $x$, $T_{eff}$, called {\it effective~temperature},
gives an estimation of the noise.
If a signal of energy $E$ due to an impulsive force acting on the bar
is generated at time $t_o$, the change of the filtered
data energy with time, neglecting the noise contribution,
 has an envelope which depends on the detector bandwidth as follows:
\be
E_s(t)=E \,\cdot \exp\left[{-2\pi\, |t-t_o|\, \Delta f} \right]
\label{segnale}
\ee
where the bandwidth $\Delta f$ is given in Table \ref{dire}.
In addition to the well behaved and modeled noise (electronic and
thermal noise), other sources of noise are active, sometimes of unknown
origin. This requires more than one detector to be simultaneously used,
in order to discriminate a real signal from noise.

For our detectors, the relationship between burst energy $E$ expressed in kelvin 
and dimensionless amplitude $h$ is given by \cite{australia}
\be
E=\frac{h^2}
{(7.97\times10^{-18})^2}
\left( \frac{\tau_{gw}}{1~{\mbox ms}}\right)^2~~~~~~~~{\mbox{[K]}}
\label{aus}
\ee
where $\tau_{gw}$ is the duration of the burst, conventionally
assumed to be $\tau_{gw}=0.001\,$s.

In the following, for a given signal of energy $E$, we shall
use the signal-to-noise ratio
\be
SNR=\frac{E}{T_{eff}}
\label{rapporto}
\ee

For the GRB's we consider the observations made with the
Gamma Ray Burst Monitor (GRBM, \cite{Frontera97}) aboard the BeppoSAX
satellite, the only satellite for the GRB detection in
operation during 2001. The GRBM is an all sky monitor which operates
in the range from 40 to 700 keV, with a GRB detection rate of about 0.7
events/day. For each GRB which triggers the on-board trigger logic,
high time resolution (up to $0.5 \,$ ms) time profiles are transmitted.
In addition, $1\,$s count rates from the GRBM in two energy 
bands ($40-700\,{\mbox{keV}}$ and $>\,100\,{\mbox{keV}}$) 
are continuously recorded and transmitted. From the GRBM data
rough information on the GRB direction can also be derived \cite{guidorzi}.

\begin{figure}
 \vspace{9.0cm}
\includegraphics{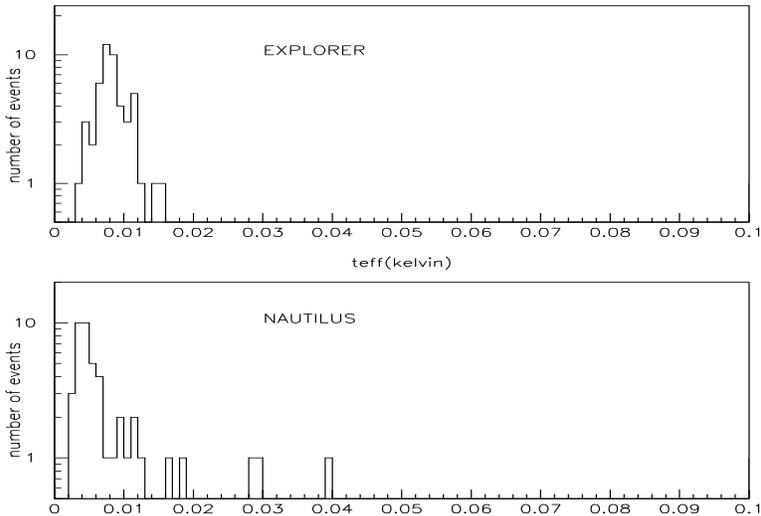}
 \caption{Distribution of $T_{eff}$ [K], in the period
 1 March through 17 July 2001.
 \label{distri_teff} }
\end{figure}

For the present analysis we use two quantities from the GRBM data:
the initial time of each GRB and its burst duration. 
In the period 1 March through 17 July 2001 we have 101 BeppoSAX bursts,
but only 51 occur at times when both antennas were operating. 
The GRB fall into two categories: one of the 38 GRB events
with short duration ($\leq \,1\,$s) and another one of 63 GRB with longer duration. The short duration events include 
all the GRB's which did not trigger the on-board logic, thus their exact
duration is not available.
%
%
Indicating with $t_\gamma$ (trigger time)
the initial time for the GRB's, for each burst we considered
the EXPLORER and NAUTILUS data in $t_m=800$ s intervals
centered at the $t_\gamma$ ($t_\gamma \, \pm\, 400$ s).
Each interval is covered  by  $800/0.004544=176056$
 data samples.
The average value of the data gives, in absence of signal,
noise temperature $T_{eff}$.
The distributions of $T_{eff}$ obtained for these stretches
of GW data are shown in Fig.~\ref{distri_teff}.
We note the good behavior of both detectors.
In view of this good behaviour we thought
to limit our analysis to the 47 stretches with $T_{eff}<20\,$mK
on both detectors, losing only four stretches.
For these stretches we have the following average values:
$\bar{T}_{eff}^{expl}=8.8\,$mK and
$\bar{T}_{eff}^{naut}=6.1\,$mK.

\section{Cross-correlation analysis}
\noindent
Due to the unknown time gap between GW's and GRB's,
it is essential to have  at least two independent GW detectors at disposal.
In such a case we can cross-correlate the two outputs, since we expect
that, if any GW burst arrives on the Earth, both detectors respond
at the same time within the travel time difference of the GW burst.
EXPLORER and NAUTILUS  being about $700\,$km apart, the
maximum travel time difference is about $3\,$ms, shorter
than the time uncertainty of the measurements, of the order of 1 sampling time. 
Since we use resonant transducers, the signal is, for each
detector, distributed over the two resonance modes, with a beat
 period of $64\,$ ms.
To cope with this problem we averaged our data over 16 samples,
that is over a time of $16\times0.004544=0.0727\,$s. Thus during each
 800 s interval we have 11004 pairs of data which we can cross-correlate.

The correlation function is defined by
\be
r(\tau)=
\frac{
\sum_i(x(t_i+\tau)-\bar{x})~(y(t_i)-\bar{y})
}
{\sqrt{\sum_i(x(t_i)-\bar{x})^2~\sum_i(y(t_i)-\bar{y})^2}}
\label{corre}
\ee
where the summations are extended over the 11004 pairs, $x(t_i)$ refers to the
EXPLORER squared data and $y(t_i)$ is the same quantity
for NAUTILUS.
$r(\tau)$ is dimensionless, by definition.
If simultaneous signals due to GW bursts arise both in the EXPLORER 
and NAUTILUS detectors,
no matter when, with respect to the GRB arrival time,
but within the time window of $\pm400\,$s, we should
find a larger value of $r(\tau)$ for $\tau=0$\,s.

\begin{figure}
 \vspace{9.0cm}
\includegraphics{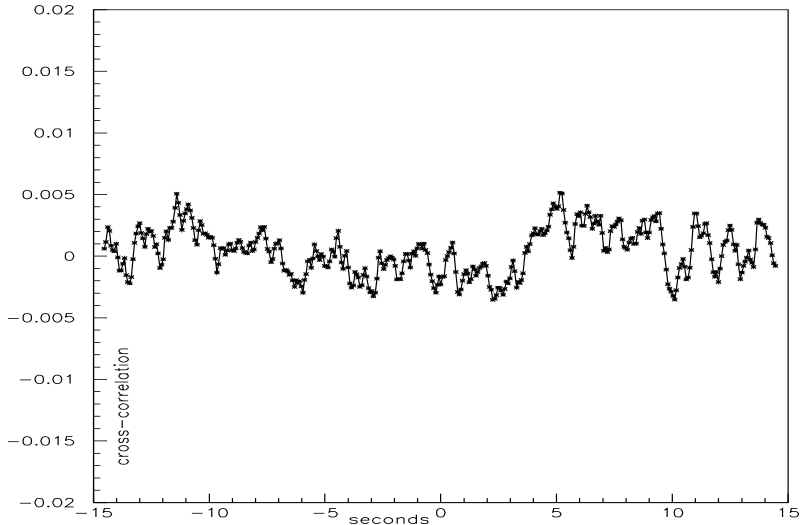}
 \caption{The cross-correlation $R(\tau)$, averaged over the 47
 GRB's versus the time shift $\tau$ in seconds. 
No positive correlation at $\tau = 0\,$s is visible.
\label{cross} 
}
\end{figure}

The 47 values of $r(\tau)$, one for each  GRB, are used to calculate
the average cross-correlation $R(\tau)$, 
shown in Fig. \ref{cross}.
No significant correlation is visible. In particular,
at $\tau=0\,$s we get the negative value of $ -0.0027$. 
We also repeated the above calculations separately for the GRB's
with short ($\leq 1\,$s) and long ($> 1\,$s) time duration.
Again, no correlation appears between EXPLORER and NAUTILUS using
a window of $\pm 400\,$s centered at the GRB arrival times.

\subsection{Modeling the average cross-correlation}
\noindent
In order to provide quantitative information out of this 
null result, we need to model our expectations for the average 
cross-correlation at zero delay time $R(0)$, hereafter
indicated with $R_0$, under the hypothesis that such signals do exist.

This modeling means going through the following considerations.

\begin{figure}
 \vspace{9.0cm}
\includegraphics{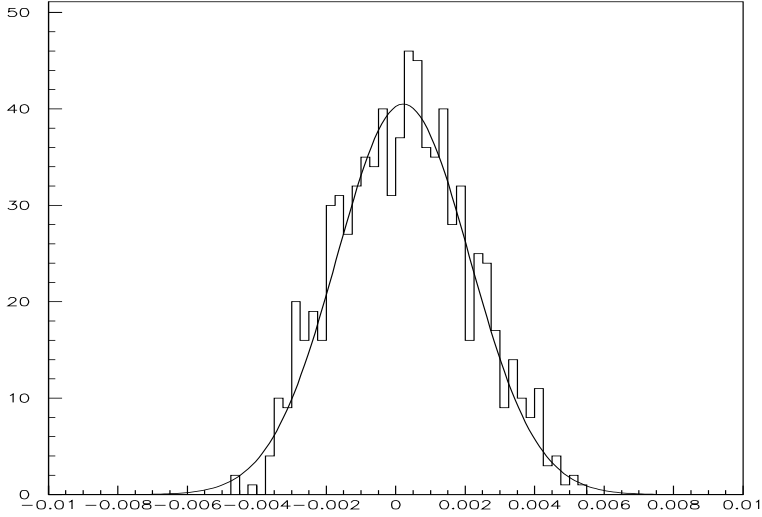}
 \caption{Distribution of the average cross-correlation $R(\tau)$ shown in Fig. \ref{cross} and Gaussian
fit. The standard deviation calculated directly from the histogram coincides with that obtained by
a Gaussian fit (0.00189 and 0.00191, respectively) }
\label{gauss}
\end{figure}

Let us suppose that GW generate signals in our two
detectors with given signal-to-noise ratios, say {\it SNR}$_{expl}$ and
{\it SNR}$_{naut}$. In the present case, since both detectors have about the
same sensitivity ($\bar{T}_{eff}^{expl}=8.8\,$mK and
$\bar{T}_{eff}^{naut}=6.1\,$mK)
and are parallel, we take, roughly, $\mbox{\it SNR}_{expl}\approx
\mbox{SNR}_{naut}\approx \mbox{SNR}$.
We define
\be
\mbox{SNR}_R=\frac{R_0}{\sigma_R}
\label{snrc1}
\ee
where $\sigma_R$ is the standard deviation of our
expectations of the dimensionless $R_0$.
This quantity can be estimated from the spread of the
many available  measurements of  $R(\tau)$ for different 
 $\tau$ (see Fig. \ref{cross}).  The distribution of $R(\tau)$ for 
all delay times, shown in Fig. \ref{gauss}, 
can be modeled by a Gaussian (see solid line in the figure), and the
standard deviation calculated directly from the histogram coincides with that obtained by
a Gaussian fit (0.00189 and 0.00191, respectively).

It is possible to evaluate  $\sigma_R$ in an alternative way (see Appendix A) ,
 using  the number of GRB's $N_{\gamma}$ and the number
of independent GW data samples $N_s$, 
in each data stretch used for calculating the
cross-correlation.  The following formula is obtained:
\be
\sigma_R=\frac{1}{\sqrt{N_s\, N_{\gamma}}}\,.
\label{sigmar}
\ee
 We estimate the number of independent
data points by the number  of independent data
of EXPLORER (the bandwidth of NAUTILUS is much narrower, see Table \ref{dire}) 
as follows
\be
N_s=t_m\,\Delta f_{expl}\approx 7200 \, ,
\label{ndata}
\ee
since the detector data are correlated within a time $1/\Delta f$.
We finally get
$\sigma_R=0.0017$,
in excellent agreement with the value 0.0019 deduced from the distribution
of $R(\tau)$, considering the roughness of our estimation of the number of 
independent data, $N_s$.

\begin{figure}
 \vspace{9.0cm}
\includegraphics{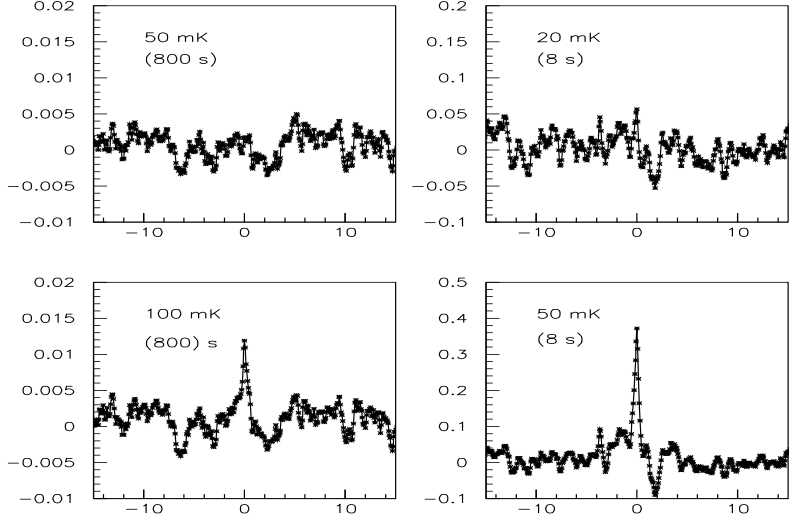}
 \caption{The dimensionless cross-correlation $R(\tau)$ averaged over the 47
 data stretches, for various applied signals.
 In the figures on the left two signals, 100 mK and 50 mK, were applied,
 under the assumption that the GW arrival times be within $\pm400\,$s
 of the corresponding GRB arrival times. The figures on the right
 show the result for applied signals of 50 mK and 20 mK, within a
  time window of $\pm4\,$s (that is, GW bursts and GRB's nearly simultaneous).
 \label{cross_doppio} }
\end{figure}

In Appendix A, the following  important relationship 
between the measured quantity ${\mbox SNR_R}$ (at $\tau=0\,$s) and the possible
signals expressed in term of signal to noise ratio {\it SNR} is obtained:
\be
\mbox{\it SNR}_R=\mbox{\it SNR}^2\,\sqrt{\frac{N_{\gamma}}{N_s}}
\label{snrc}
\ee 
In order to check the model for the average cross-correlation at zero
delay time, and thus Eq. \ref{snrc}, we performed a test by adding to
the data, for each GRB, signals of given amplitude at the
same time for the two detectors.
The results are given in Fig.\ref{cross_doppio}, where we show
the cross-correlation for some applied signals
for the two time windows of $\pm400\,$s and $\pm4\,$s.
The results here are given in terms of the dimensionless average cross-correlation,
but in the figure we have also indicated  the energies of the input signals.
We notice that, as predicted by Eq. \ref{snrc}, the ${\mbox SNR_R}$ 
increases by reducing the intervals of the cross-correlation, for a given ${\mbox SNR}$.

\subsection{Relations between $\mbox{SNR}_R$, burst energy and dimensionless amplitude}
\noindent
Since we require the result in terms
of GW amplitudes $h$, we need to relate cross-correlation quantities
to energies. 

From Eqs.~(\ref{rapporto}) and (\ref{snrc}) we obtain
\be
E=T_{eff} \left( \frac{ \mbox{\it SNR}_R^2 N_s}{N_{\gamma}}\right)^{1/4}
\hspace{0.6cm}\mbox{[K]}
\label{upperE}
\ee
It would appear that $E$ decreases with increasing $N_s$, i.e. bandwidth, but 
it can be demonstrated that $T_{eff}$ decreases linearly with increasing bandwidth. 

In terms of the dimensionless amplitude, using Eq.~(\ref{aus}) 
with $\tau_{gw}=\,$1 ms we get,
\be
h=7.97\times10^{-18}\sqrt{E}
\label{upperh}
\ee
with E expressed in kelvin.
We finally get, using Eqs.~(\ref{snrc1}) and (\ref{upperE})
\be
\mbox{SNR}_R=\left(\frac{E}{T_{eff}}\right)^2 \cdot \sqrt{\frac{N_\gamma}{N_s}}
\label{eq:sigmar_e}
\ee
We observe that a negative value of $\mbox{SNR}_R$ gives a negative value for
$E^2$. 
To check the validity of Eq.~(\ref{upperE}), we applied several
signals for various time windows of correlation and compared the energies of the input
simulated signals to the values calculated using  Eq.~(\ref{upperE}). 
This is done in the following way: we note that $R_0$, in the absence 
of applied signals, is not null for each time window.
We determine the signal we must apply in order to increase the
value of $R_0$, obtained before the application of the signals,
 by two standard deviations. We then use Eq.(\ref{upperE})
with the value $SNR_R=2$.
The result is shown in Table \ref{finale}.
\begin{table}
\centering
\caption{
Comparison of energies calculated, using Eq.~(\ref{upperE}),
and simulated (values of an input signal that produces $SNR_R$=2).
}
\vskip 0.1 in
\begin{tabular}{|c|c|c|}
\hline
window&$E$ [mK]&$E$ [mK]\\
s&calculated, Eq.~(\ref{upperE}) &simulated (input signals)\\
\hline
$\pm 400$&43.8&45\\
$\pm 40$&24.6&27\\
$\pm 16$&19.6&19\\
$\pm 4$&13.8&14\\
\hline
\end{tabular}
\label{finale}
\end{table}
The agreement between the values of the simulated input signals and 
the values calculated using Eq.~(\ref{upperE}) shows that our model is correct.

\section{Inference of the GW burst amplitude}
\noindent
Having presented the experimental method and the model
for the averaged correlation at zero delay time $R_0$,
we can infer the values of GW amplitude $h$
consistent with the observation.
We note that, using Eqs. \ref{upperE} and \ref{snrc1}, energy $E_0$ is
related to the measured cross-correlation $R_0$ by 
\be
E_0=T_{eff} \left( \frac{N_s}{N_{\gamma}}\right)^{1/4} \cdot
\left( \frac{R_0} {\sigma_R}\right)^{1/2}
\hspace{0.6cm}\mbox{[K]}
\label{eq:relazione}
\ee
Hence, the data are summarized by an observed average squared energy 
$E_0^2 = - 1.11 \times 10^{-3}\,\mbox{K}^2$,
at $-1.4$ standard deviation from the expected
value in the case of noise alone, as calculated with the aid of Eq. (\ref{eq:sigmar_e}) where we put
$\mbox{SNR}_R =- 1.4$.
The standard deviation, expressed in terms of squared
energy, is obtained from Eq. (\ref{eq:sigmar_e}), in the case $\mbox{SNR}_R=1$,
which gives  $\sigma_{E^2} = 0.79\times 10^{-3}\,\mbox{K}^2$.

\subsection{Probabilistic result and upper limits}
\noindent
According to the model discussed above, in the case of
GW signals of energy $E$, we expect $E_0^2$ to be
a random number, modeled with a Gaussian probability
density function around $E^2$ with a standard
deviation $\sigma_{E^2}$:
\be
 f(E_0^2\,|\,E^2) \propto 
\exp{\left[-\frac{(E_0^2-E^2)^2}{2 \sigma_{E^2}^2}\right]}\,,
\ee
where $E$ is the unknown quantity we wish to infer from
the observed value of $E_0$, given in Eq. \ref{eq:relazione}.
This probability inversion is obtained using Bayes' theorem
(see, e.g., \cite{giuliocern} for a physics oriented introduction):
\be
f(E^2\,|\,E_0^2) 
\propto f(E_0^2\,|\,E^2)
\cdot f_\circ(E^2)\,
\label{eq:Bayes}
\ee
where $ f_\circ(E^2)$ is the prior probability density
function of observing GW signals of squared energy $E^2$.
In fact, we are eventually interested in inferring the GW's amplitude $h$, related 
to the energy $E$ by Eq. (\ref{upperh}).
Therefore we have a similar equation:
\be
f(h\,|\,E_0^2) 
\propto f(E_0^2\,| \,h)
\cdot f_\circ(h)\,
\label{eq:Bayesh}
\ee
where $ f(E_0^2\,|\,h)$ is obtained by a 
transformation of  $ f(E_0^2\,|\,E^2)$.
As prior for $h$ we considered a uniform distribution, 
bounded to non negative values
of $h$, obtained from Eq. \ref{eq:Bayesh}, i.e. $f_\circ(h)$ 
is a step function $\theta(h)$.
This seems to us a reasonable choice and it is stable, 
as long as other priors can be conceived which
model the ``positive attitude of reasonable scientists`` 
(see Ref. \cite{giuliocern,pia})
\footnote{A prior distribution alternative to the uniform can be based
on the observation that what often seems uniform is not the probability
per unit of $h$, but rather the probability per decade of $h$, 
i.e. researchers may feel equally uncertain about 
the orders of magnitudes of $h$. This prior is known as  Jeffreys' prior, but, in our case, 
it produces a divergence for $h\rightarrow 0$ in Eq. \ref{eq:Bayesh}, a
direct consequence of the infinite orders of magnitudes which are equally
believed.
To get a finite result we need to put a cut-off at a given value of $h$. 
This problem is described in depth, 
for example, in \cite{pia} and in \cite{priors}. \label{foot:1}}.

\begin{figure}[t]
\begin{center}
\epsfig{file=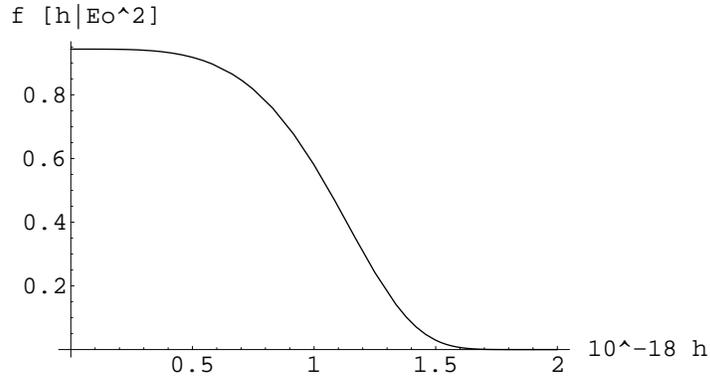,clip=,width=0.6\linewidth} 
 \end{center}
\caption{Probability density function   
$f(h\,|\, E_0^2)$ (see Eq. \ref{eq:Bayesh}). 
The prior used for this calculation is a step function. }
\label{finaleh} 
\end{figure}

The probability density function of $h$ is plotted in Fig. \ref{finaleh}.
The highest beliefs are for very small values, while values above 
$1.5\times 10^{-18}$ are practically ruled out. 
From Fig. \ref{finaleh} we obtain an expected value and standard deviation
for $h$ of $0.56\times 10^{-18}$ and $0.35\times 10^{-18}$, 
respectively, which fully account for what is perceived as a null result.

\begin{figure}[t]
\begin{center}
\epsfig{file=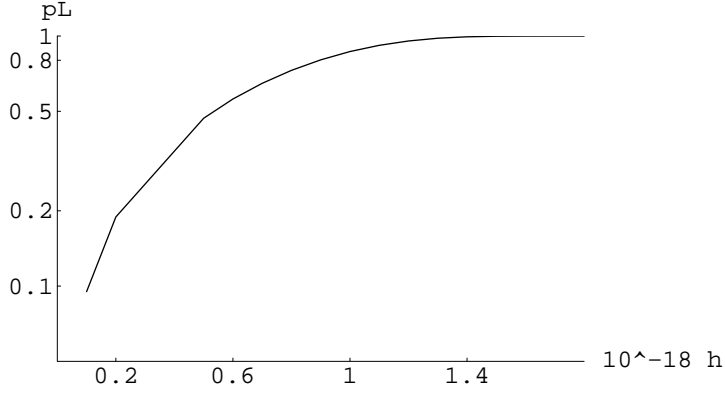,clip=,width=0.6\linewidth} 
\end{center}
\caption{Upper limits for the amplitudes $h$. Y-axis gives the probability 
$p_L(h) \le h$.}
\label{fighh} 
\end{figure}

In these circumstances, we can provide an upper limit,
defined as value $h(UL)$, such that there is a given probability for 
the amplitude of GW's to be below it, i.e.
\be
\int_0^{h(UL)}\!\!f(h \,|\,E_0^2)\,\mbox{d}h = p_L\,,
\ee
with $p_L$ the chosen probability level. 
Results are plotted in  Fig.\ref{fighh}.
For example, we can exclude the presence of
signals of amplitudes $h \ge 1.2 \cdot 10^{-18}$ with 95\% probability.

\subsection{Prior-independent result and upper sensitivity bound}
\noindent
Probabilistic results depend necessarily on the choice of prior
 probability density function of $h$. For example, those firmly convinced
that GW burst intensities should be in the $10^{-22}$ region
 would never allow a 5\% chance to
$h$  above  $1.2 \cdot 10^{-18}$.
Therefore, in frontier research particular care has to be used, before 
stating probabilistic results.
The Bayesian approach, thanks to the factorization between 
likelihood and prior, 
offers natural ways to a present prior-independent 
result. The simple idea would be just to provide
 the likelihood for each hypothesis under investigation,
in our case  $f(E_0^2\,|\, h)$. 
More conveniently, it has been proposed 
(\cite{ZEUS}-\cite{pia}) to publish 
 the likelihood rescaled to the asymptotic
limit, where experimental sensitivity is lost completely;
$h = 0$, in our case. Indicating with ${\cal R}$ 
this rescaled likelihood, we have 
\be
{\cal R}(h) = \frac{f(E_0^2\,|\, h)}
                        {f(E_0^2 \,|\,h_{ref}=0)}\,.
\label{eq:rbur_def}
\ee
In statistics jargon, this function gives the Bayes factor of all
$h$ hypotheses with respect to $h=0$. In intuitive
terms, it can be interpreted as a ``relative belief updating ratio''
or a `` probability density function shape distortion function'', 
since from Eq.~(\ref{eq:Bayesh}) we have
\be
\frac{f(h\,|\,E_0^2) }
{f(h=0\,|\,E_0^2) } =
\frac{f(E_0^2\,|\, h)}
                 {f(E_0^2\,|\, h=0)}
           \,\cdot\,
\frac{f_\circ(h)}{f_\circ(h=0)}\,.
\label{eq:Bayesf}
\ee
In the present case we get, numerically
\be
{\cal R}(h) = A \cdot \exp\left[-\frac{(B-h^4)^2}{C} \right] \hspace{0.5cm} 
  \mbox{for }h \ge 0\,, 
  \ee
where $A = 2.6645$ is the rescaling factor, 
$B = -4.4 \times \left(10^{-18}\right)^4$ and 
$C=20.4\times \left(10^{-18}\right)^4\,.$

\begin{figure}[t]
\begin{center}
\epsfig{file=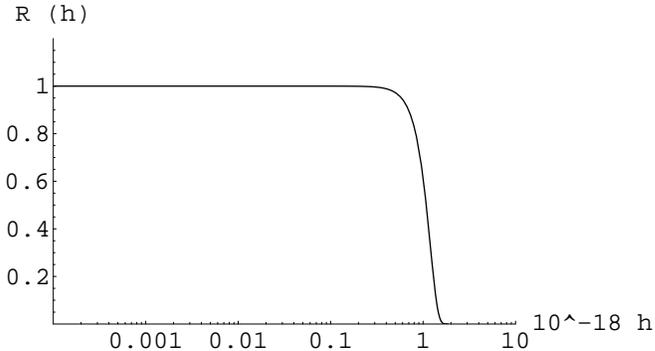,clip=,width=0.6\linewidth}
\end{center}
\caption{Relative belief updating ratio, as a function of the 
dimensionless amplitude of GW's, plotted in log-linear scale.}
\label{loglinRh}
\end{figure}

The result is given in Fig.~\ref{loglinRh},
where the choice of the log scale for $h$ 
is to remember that there are infinite orders of magnitudes
where the value could be located (and hence the problem discussed in footnote
\ref{foot:1}). 
Interpretation of Fig.~\ref{loglinRh}, in the light
of Eqs.~(\ref{eq:rbur_def})-(\ref{eq:Bayesf}), is straightforward:
up to a fraction of $10^{-18}$ the experimental evidence does not
produce any change in our belief, while values much larger 
than   $10^{-18}$ are completely ruled out. The region of transition
from ${\cal R}$ from 1 to zero identifies a {\it sensitivity bound} for 
the experiment. The exact value of this bound is a matter of convention,
and could be, for example, at ${\cal R} = 0.5$, or ${\cal R} = 0.05$. 
We have $1.3 \times 10^{-18}$ and $1.5 \times 10^{-18}$, 
respectively. Note that these bounds have
no probabilistic meaning. In any case,
the full result should be considered to be the  ${\cal R}$ 
function, which, being proportional to the likelihood,
can easily be used to combine results 
(for independent datasets the global likelihood is 
the product of the likelihoods, and proportional constants
can be included in the normalization factor).
Note that the result given in terms of scaled likelihood and sensitivity bound 
cannot be misleading. In fact, these results are not probabilistic 
statements about $h$ and no one would imagine they were. On the
other hand, ``confidence limits', which are not probabilistic
statements on the quantity of interest, tend to be perceived as 
such (see e.g. \cite{GiulioME98} and references therein).
  

\section{Conclusion}
\noindent
Using for the first time a cross-correlation
method applied to the data of two GW detectors, EXPLORER and NAUTILUS,
new experimental upper limits have been determined for
the burst intensity causing correlations of GW's with GRB's. 
Analyzing the data over 47 GRBs, we exclude the presence
of signals of amplitude $h_{GW}\ge 1.2 \cdot 10^{-18}$, with 95\% probability,
with a time window of $\pm~400~s$.
With the time window of $\pm~5$ s, we improve the previous
GW upper limit to about $h=6.5~10^{-19}$.

The result is also given in terms of scaled likelihood and sensitivity
bound, which we consider the most complete and unbiased way 
of providing the experimental information.

In a previous paper \cite{vulcano} we had given more stringent upper
limits, but this was
under the hypothesis that the GW signals always occur at the same time with respect
to the GRB arrival time. Here, instead, we only require that the time gap
between the GRB and the GW burst be within a given time window.
Similar comparison can be made with the AURIGA/BATSE 
result \cite{cerdonio}, where an 
upper limit `` $h_{RMS} \le 1.5~10^{-18}$ with C.L. 95\% ''
is estimated under the assumption that GW's arrive
at the GRB time within a time window of $\pm5$ s. 

Finally, we remark that this method can be applied for any expected delay
between GRB and GW, with appropriate time shifting of the integration
window with respect to the GBR arrival time,
according to the prediction of the chosen model.
\section{Acknowledgments}
\noindent
We thank F. Campolungo, R. Lenci, G. Martinelli, E. Serrani ,
 R. Simonetti and F. Tabacchioni for precious technical assistance.
We also thank Dr. R. Elia for her contribution.

\section*{Appendix A}
Given two independent detectors, let $x(t)$ and $y(t)$ be the measured quantities,
which are the  (filtered) data in our case. 

We introduce the variables $\xi(t)=x(t)-\bar{x}$
and $\eta(t)=y(t)-\bar{y}$ where $\bar{x}=E[x]=T_{eff}$
and $\bar{y}=E[y]=T_{eff}$. We recall that the cross-correlation function is
\be
r(\tau)=\frac{\sum_i (\xi(t_i) \cdot \eta(t_i+\tau))}
{\sqrt{\sum_i \xi^2  \sum \eta^2}}
\label{crossa}
\ee
with summation extended up to the number of independent samples $N_s$.
We calculate $\sum\xi^2=\sum\eta^2=N_sT_{eff}^2$. Thus
$\sqrt{\sum\xi^2\cdot\sum\eta^2}=N_sT_{eff}^2$.
We easily verify that $E[r(\tau)]=0$. In the absence of any correlated signal
we also have $E[r(0)]=0$.

Let us calculate the variance of $r(\tau)$. We notice that when
squaring the numerator of Eq.\ref{crossa} and taking the average, the cross-terms
vanish if the $N_s$ data are independent from each other. Then,
since also $\xi$ and $\eta$ are independent variables, we obtain

\be
\sigma_r^2=\frac{\sum_i(\xi(t_i)\eta(t_i+\tau))^2}{(N_s T_{eff}^2)^2}=
\frac{N_s\cdot E[\xi^2]E[\eta^2]}{(N_sT_{eff}^2)^2}=\frac{1}{N_s}
\label{siga}
\ee

The previous considerations still apply to the cumulative cross-correlation
$R(\tau)$, obtained by averaging $N_{\gamma}$ independent $r(\tau)$.
The final variance for the cross-correlation $R(\tau)$ is
\be
\sigma_R^2=\frac{1}{N_sN_{\gamma}}
\label{sigma}
\ee

Let us now consider an energy signal $S$ on both detectors at the same time.
The expected value of the cross-correlation $r(\tau)$ at $\tau=0$ will be positive.
In the case of $N_s$ independent data the signal $S$ will appear
in one datum only and we have:
\be
E[R(\tau=0)-R(\tau\neq 0)]=\frac{S^2}{N_sT_{eff}^2}=\frac{SNR^2}{N_s}
\label{segnalea}
\ee
where we put $SNR=\frac{S}{T_{eff}}$.
We note that, in Eq.\ref{segnalea},  $E[R(\tau\neq0)]=0$ also when a signal
is present.

 Putting also
\be
\mbox{SNR}_R=\frac{E[R(\tau=0)]}{\sigma_R}
\label{snrr}
\ee
(as already been defined in the text in Eq. \ref{snrc1})
we obtain
\be
\mbox{SNR}_R=\mbox{SNR}^2\sqrt{\frac{N_{\gamma}}{N_s}}.
\ee

\end{document}